\documentclass[pra,showpacs,preprintnumbers,amsmath,amssymb,twocolumn]{revtex4}
\usepackage{bm}
\usepackage{dcolumn}
\usepackage{graphicx}
\usepackage{amsbsy}
\usepackage{amsmath}
\usepackage{amsfonts}
\usepackage{amsthm}
\usepackage{color}
\usepackage{mathrsfs}
\usepackage{yfonts}

\begin{document}
\hyphenation{act-ually angle local-iz-ation local-ized micro-wave momen-tum numer-ical numer-ically para-meter para-meters}

\theoremstyle{plain}
\newtheorem{theorem}{Theorem}
\newtheorem{lemma}[theorem]{Lemma}
\newtheorem{corollary}[theorem]{Corollary}
\newtheorem{conjecture}[theorem]{Conjecture}
\newtheorem{proposition}[theorem]{Proposition}

\theoremstyle{definition}
\newtheorem{definition}{Definition}

\theoremstyle{remark}
\newtheorem*{remark}{Remark}
\newtheorem{example}{Example}

\title{Quantum Localization for Two Coupled Kicked Rotors}%
\author{Borzumehr Toloui}%
\email{btoloui@qis.ucalgary.ca}
\affiliation{Institute for Quantum Information Science, University of Calgary, Calgary, Alberta, Canada T2N 1N4.\\
Physics Department, Simon Fraser University, Burnaby, British Columbia, Canada V5A 1S6.}
\author{Leslie E. Ballentine}%
\email{ballenti@sfu.ca}
\affiliation{Physics Department, Simon Fraser University, Burnaby, British Columbia, Canada V5A 1S6.}
\date{\today}%
\begin{abstract}
We study a system of two coupled kicked rotors, both classically and quantum mechanically,
for a wide range of coupling parameters.  This was motivated by two published reports, one
of which reported quantum localization, while the other reported diffusion.
The classical systems are chaotic, and exhibit normal diffusive behavior.
In the quantum systems, we found different regimes, depending on the strength of the coupling.
For weak coupling, we found quantum localization similar to that exhibited by single kicked rotors. 
For strong coupling, we found a quasi-diffusive growth of the width of the momentum distribution,
in which the apparent diffusion coefficient decreased as time increased. 
The behavior in this regime can be described by the scaling theory of weak localization for
two-dimensional disordered lattices. 
The weak and strong coupling regimes are separated by a regime of complex intermediate behavior. 
Thus we explain the apparent contradiction in the literature as being due to qualitatively
different regimes of behavior, which we call strong and weak quantum localization.
\end{abstract}
\maketitle

\section{Introduction}
Kicked rotor models~\cite{kr-gen1,kr-gen2,kr-gen3}  have played a prominent role in the study of classically chaotic systems and their corresponding quantum behavior.
Although quite simple, these models exhibit most of the distinctive features
of classical and quantum chaos.
The particular phenomenon that is the subject of this paper is {\it quantum localization},
which is a localization in angular momentum space that is in sharp contrast to the
diffusive motion typical of classical chaos.
The understanding of quantum localization is considerably facilitiated by the existence
of a mapping from the rotor model to a model of a particle moving in a disordered lattice~\cite{Fishman},
which exhibits the phenomenon of Anderson localization~\cite{Anderson}.

Single rotor systems have been thoroughly studied and are well understood~\cite{Fishman}.
However, the situation is less clear for systems of two coupled kicked rotors.
Indeed, there are examples in the literature whose results appear to
be contradictory.
Doron and Fishman~\cite{Doron} have studied one such model, and
they found quantum localization of the Anderson type, scaling exponentially with the
coupling strength. 
On the other hand,  Adachi {\it et~ al.}~\cite{Ikeda1} have studied a model with 
different coupling potentials, and have obtained results that exhibit diffusive growth in the
width of the state.
They concluded that the coupling between the two rotors can restore the mixing \cite{Ikeda2} 
that would otherwise be suppressed by quantum localization. 

There are several possible explanations for this apparent contradiction.
The two papers are based on models that differ in the form of the kicking and coupling potentials.
Although it is logically possible for the two models to behave qualitatively differently,
this would be very surprizing, since the forms of the potentials are similar enough that
we should expect them to belong to the same generic class.
The ranges of interaction parameter strength used in the two papers are quite different.
Indeed, the lattice emerging from the choice of parameters in~\cite{Ikeda1} turns out to
be periodic, which would give rise to extended Bloch states.  So we need to study the 
two models in equivalent ranges of paramenters.
The two papers also used different criteria for identifying localization.
Doron and Fishman~\cite{Doron} look for an exponential fall-off in the angular momentum 
distribution, while Adachi {\it et~ al.}~\cite{Ikeda1} look for a saturation in time of
the width (standard deviation) of the distribution.
So we need to apply both criteria to both models to determine whether they agree.
A further complication arises because two-dimensional disordered lattices contain 
the marginal case of weakly localized states, which creates more interesting possibilities.

In this paper we investigate these questions.  
To do so, we consider a more general Hamiltonian that includes the models of~\cite{Doron} and~\cite{Ikeda1}  as special cases. 
We study the system numerically for a suitable range of parameters,
and examine the shape and fall-off of the angular momentum distribution,
together with its standard deviation as a measure of the state width.  
In Section II we discuss the Hamiltonian, the initial state, and the method of numerical 
calculation.  In Section III the classical behavior of the system is studied, and is shown 
to lie within the chaotic region of phase space for the chosen range of parameters. 
Section IV presents the numerical results for the quantum system. 
In Section V we examine whether the results are consistent with the scaling properties of 
weak localization. 
Section VI discusses the conclusions.   
\section{Model}

The general Hamiltonian is chosen to be
\begin{align}
&H=T(p_{1}, p_{2}) +  V(\theta_{1}, \theta_{2})  \sum_{n=-\infty}^{+\infty} \delta(t-n)& \nonumber \\
&T(p_{1}, p_{2})=\frac{1}{2} \;\alpha_{1} \;p_{1}^{2}+\frac{1}{2} \;\alpha_{2} \;p_{2}^{2}&\nonumber
\end{align}
\begin{align}
V\left(\theta_{1}, \theta_{2}\right)=[\lambda_{1}  \cos \theta_{1} +\lambda_{2}  \cos \theta_{2}+ \lambda_{3}  \cos \theta_{1}\;\cos \theta_{1} \nonumber \\ + \lambda_{4} \cos ( \theta_{1}-\theta_{2} )]
\end{align}
The dimensionless parameters are related to the physical quantities as follows:
\begin{equation}
\label{alphas}
\alpha_{1} = \frac{\hbar \tau}{I_{1}}, \:\:\:\alpha_{2} = \frac{\hbar \tau}{I_{2}}
\end{equation}
Here $\tau$ is the time span between kicks, and $I_{1}$, $I_{2}$ are the moments of inertia of 
the first and second rotors, respectively. 
The parameters $\lambda_{1}$ and $\lambda_{2}$ are the single-rotor kicking strengths, 
while $\lambda_{3}$ and $\lambda_{4}$ are the couplings between the two rotors.  

The choice  $\lambda_{1} = \lambda_{2}=\lambda_{4}=0.0$ and $\alpha_{1}=1,
 \alpha_{2}=\sqrt{2},\, \hbar=1$ will yield the Hamiltonian studied in~\cite{Doron}. 
The main systems analyzed in~\cite{Ikeda1} correspond to 
 $\lambda_{3}=0.0$, $\alpha_{1}=\alpha_{2}=\frac{41}{512}\times 2\pi$, and $\alpha_{1}=\alpha_{2}=\frac{41}{1024}\times 2\pi$.  
Since these values of $\alpha$ are commensurate with $\pi$, they yield periodic lattices,
giving rise to extended Bloch-like states and excluding the possibility of Anderson localization. 
Therefore, we have used a different value in our numerical studies.

The quantum system is obtained by replacing the momentum variables with operators
  $\hat{p}_{1}= -\imath \partial /\partial \theta_{1}$ and $\hat{p}_{2}= -\imath \partial /\partial \theta_{2}$. 
The time evolution for one period of the kicking Hamiltonian is given by the Floquet operator
\begin{eqnarray}
\hat{U}= e^{-\imath V(\theta_{1}, \theta_{2})} \; e^{-\imath T(\hat{p}_{1}, \hat{p}_{2})}
\label{U}
\end{eqnarray}
The numerical calculation is performed in two stages. Starting in the momentum representation,
 the propagator $e^{-\imath T(\hat{p}_{1}, \hat{p}_{2})}$, which is diagonal in this basis,
 is applied for the duration between two kicks. 
Then the kicking propagator $e^{-\imath V(\theta_{1}, \theta_{2})}$ is applied in the 
angle representation. The transformation between the angle and angular momentum representations
 is achieved by Fast Fourier Transform (FFT) and its inverse.  
The same process is then repeated for the next period, and so on. 

The initial state is chosen to have the form
\begin{align}
\Psi(\theta_{1}, \theta_{2}) = \psi(\theta_{1}) \; \psi(\theta_{2}) \nonumber \\
\psi(\theta) = \sum_{m} \; a_{m} \; e^{\imath m (\theta-\theta_{0})} \nonumber \\
a_{m} = e^{- [\frac{m-m_{0}}{2 \Delta m}]^{2}}
\end{align}
This is a product of Gaussian wave packets, centered at momentum $m_{0}$
and angle $\theta_{0}$, with momentum width $ \Delta m = 1.25786$.
Hence we have $\hbar \Delta m \, \Delta \theta = \hbar/2$.
A summary of the parameter values are given in Table \ref{Tablez}. 
\begin{table}
\caption{Parameters of the initial state}
\label{Tablez}
\begin{center}
\begin{tabular}{|c|c|c|c|} 
\hline
Rotor no. & $m_{0}$ & $\Delta m$ & $\theta_{0}$\\
\hline
\hline
1 & 0 & 1.25786 & 0\\
\hline
2 & 0 & 1.25786 & 0\\
\hline
\end{tabular}
\end{center}
\end{table}

The momentum values form a discrete grid. 
The angle values are also set on a grid of the same size, since the Fast Fourier Transforms are 
instances of the discrete Fourier transform algorithm. 
The grid size is chosen to be $2^{11}=2048$ grid points, ranging from $m= -1023$  to $1024$ of 
dimensionless angular momentum values. The numerical calculation is performed for 30,000 kicking steps.  
The coupling parameters $\lambda_{3}$ and $\lambda_{4}$ range from $0.0$ to $3.0$ 
in steps of $0.5$, for each value of $\lambda_{1}$ and $\lambda_{2}$ in the set 
$\left\{0.0, 0.25, 0.5\right\}$. 
The simulations are done for the values $\alpha_{1}=\alpha_{2}=1.0$, which are not commensurate 
with $\pi$, thus ensuring that the corresponding Anderson lattice is non-periodic.

\section{Classical Diffusion}

In order to analyze the quantum dynamics, it is necessary to determine whether the 
classical counterpart lies in the chaotic regions of phase space. 
Only when it is verified that the corresponding classical evolution is chaotic, can the 
localization in the quantum system be positively attributed to quantum effects. 
It is also interesting to compare the behavior of the quantum system to its classical 
counterpart. 

The classical mapping of each point $(\theta^{[n]}_{1}, \theta^{[n]}_{2}, p^{[n]}_{1}, p^{[n]}_{2})$ of the phase space from its values at kick $n$ to  kick $n+1$ is:
\begin{widetext}
\begin{align}
\begin{cases}
p^{[n+1]}_{1} = p^{[n]}_{1} + \lambda_{1} \, \sin \theta^{[n]}_{1} + \lambda_{3} \, \sin \theta^{[n]}_{1} \cos \theta^{[n]}_{2} +\,\lambda_{4} \, \sin ( \theta^{[n]}_{1}-\theta^{[n]}_{2}) \\
p^{[n+1]}_{2} =p^{[n]}_{2} + \lambda_{2} \, \sin \theta^{[n]}_{2} +\lambda_{3} \, \cos \theta^{[n]}_{1} \sin \theta^{[n]}_{2} -\,\lambda_{4} \, \sin ( \theta^{[n]}_{1}-\theta^{[n]}_{2}) \\
\theta^{[n+1]}_{1} =\theta^{[n]}_{1}+p^{[n+1]}_{1} \\
\theta^{[n+1]}_{2} =\theta^{[n]}_{2}+p^{[n+1]}_{2}
\end{cases}
\label{clasmap}
\end{align}
\end{widetext}

\begin{figure}[h]
 \centering
\includegraphics[width=3.0in]{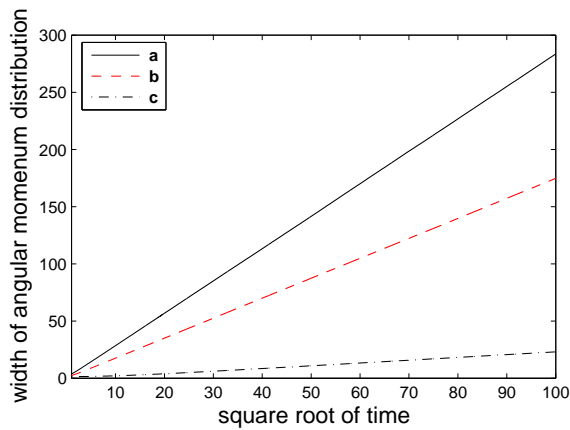}
 \caption{(Color online) Width of the angular momentum distribution for a classical rotor [defined similar to (\ref{width})]  {\bf a}: $\lambda_{1}= 0.5, \:\lambda_{2}= 0.5, \:\lambda_{3}=3.0, \: \lambda_{4}=3.0$.
{\bf b}: $\lambda_{1}= 0.5, \:\lambda_{2}= 0.5, \:\lambda_{3}= 1.0,\: \lambda_{4}=2.0$.
{\bf c}: $\lambda_{1}= 0.5, \:\lambda_{2}= 0.5, \:\lambda_{3}=1.0, \: \lambda_{4}=0.0$.}
 \label{fig:1}
 \end{figure}

For the classical calculations, an ensemble of 1,000,000 initial states was created, with the same 
Gaussian distribution in phase space as the initial quantum state. 
Each member of the ensemble was evolved separately using the mapping (\ref{clasmap}). 
The results of the calculation show similar chaotic behavior for all of our choices of parameters,
with the exception of $\lambda_1 = \lambda_2 = \lambda_3 = 0$ and $\lambda_4 > 0$, for which
the total angular momentum is a constant of motion.
The variance of the angular-momentum distribution for each rotor exhibited 
diffusive growth with time,
\begin{align}
\label{cl-dif}
\begin{cases}
\left<p_{1}^{2}\right>-\left<p_{1}\right>^{2}= D_{1}\; t\\
\left< p_{2}^{2}\right>-\left<p_{2}\right>^{2}= D_{2} \; t
\end{cases}
\end{align}
where the average is over the classical ensemble.
This is illustrated in Fig.\ref{fig:1}.
From this we conclude that the chosen parameters do lie within the classically chaotic region 
of parameter space.   

\begin{figure}[htp]
 \centering
 \includegraphics[width=3.6in]{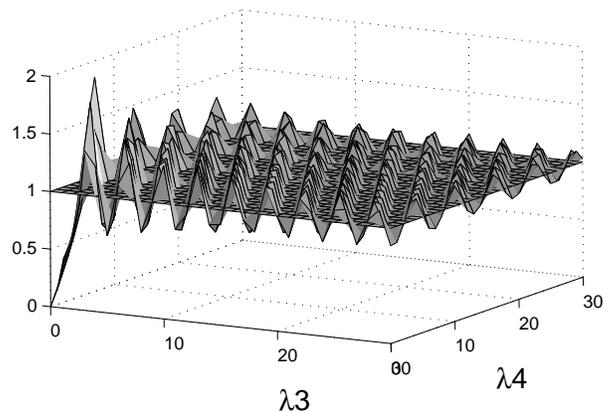}
 \caption{Ratio of the classical diffusion coef. to the quasi-linear approx.,
 $D_{1}/D^{0}_{1}$, vs couplings $\lambda_{3}$ and $\lambda_{4}$,
 for $\lambda_{1}= 0.5, \: \lambda_{2}= 0.5$. }
 \label{fig:2}
 \end{figure}

For our chosen initial states, we have $\left<p\right> = 0$ in all cases, so 
we can replace the variance with $\left<p^{2}\right>$. 
The dependence of the diffusion coefficients, $D_i$, on the kicking and coupling parameters 
can be analyzed in the lowest order (so-called {\it quasi-linear}) approximation 
by assuming a uniform probability 
distribution for the angle after a sufficiently large number of kicking steps $n$, 
and integrating  $(p^{[n+1]}-p^{[n]})^{2}$ over this distribution for each of the 
two rotors~\cite{LL}: 
\begin{align}
\begin{cases}
D^{0}_{1} =\frac{1}{2\pi} \int_{0}^{2\pi}{ \left(p^{[n+1]}_{1}-p^{[n]}_{1}\right)^{2} \; d\theta^{[n]}_{1} d\theta^{[n]}_{2}} \\ \\
D^{0}_{2} =\frac{1}{2\pi} \int_{0}^{2\pi} \left(p^{[n+1]}_{2}-p^{[n]}_{2}\right)^{2} \; d\theta^{[n]}_{1} d\theta^{[n]}_{2} \\
\end{cases}
\end{align}
This yields
\begin{align}
\begin{cases}
D^{0}_{1} = \frac{(\lambda_{3}+\lambda_{4})^2}{4} + \frac{\lambda_{4}^{2}}{4} +\frac{\lambda_{1}^{2}}{2}\\ \\
D^{0}_{2} = \frac{(\lambda_{3}+\lambda_{4})^2}{4} + \frac{\lambda_{4}^{2}}{4} +\frac{\lambda_{2}^{2}}{2}
\end{cases}
\end{align}  
A comparison between this approximation and the numerically computed diffusion constant is shown
in Fig.\ref{fig:2}.
The computed diffusion coefficients oscillate above and below the approximate value.  
This is  similar to a known result for the standard map, for which
one has $D^{0} = \frac{K^{2}}{4}$, where $K$ is the kicking strength in the standard map.  This result was also verified by a more systematic method similar to that used in~\cite{LL}.

\section{Results for the Quantum Rotors}

\begin{figure}[htp]
 \centering
 \includegraphics[width=2.9in]{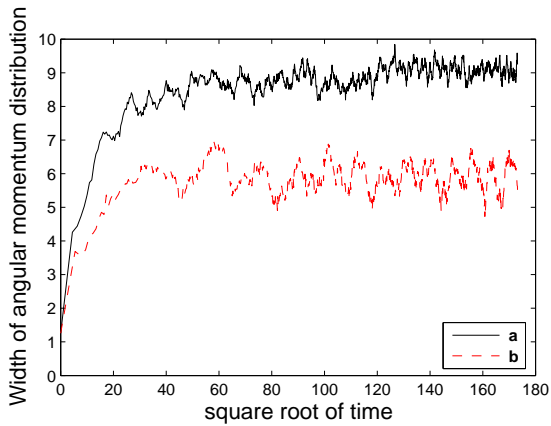}
 \caption{(Color online) Localized regime. \\ 
 {\bf a}: $\lambda_{1}= 0.5, \:\lambda_{2}= 0.5, \:\lambda_{3}=2.5, \:\lambda_{4}=0.5$. \\
 {\bf b}: $\lambda_{1}= 0.25, \:\lambda_{2}= 0.25, \:\lambda_{3}= 2.5, \:\lambda_{4}=0.5$.}
 \label{fig:3}
 \end{figure}

\begin{figure}[htp]
 \centering
 \includegraphics[width=2.9in]{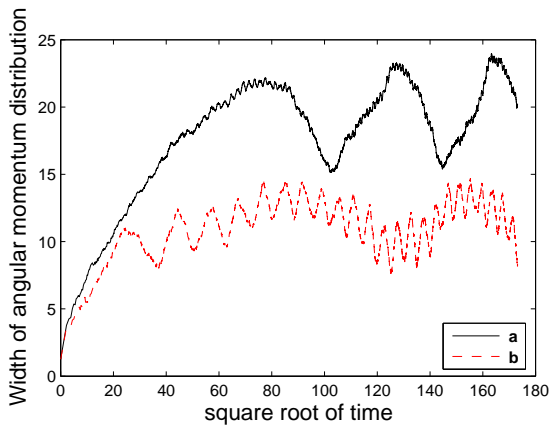}
 \caption{(Color online) Intermediate regime. {\bf a}: $\lambda_{1}= 0.5, \:\lambda_{2}= 0.5, \:\lambda_{3}=2.5, \:\lambda_{4}=1.0$.
 {\bf b}: $\lambda_{1}= 0.25, \:\lambda_{2}= 0.25, \:\lambda_{3}= 2.5, \:\lambda_{4}=1.0$.}
 \label{fig:4}
 \end{figure}

\begin{figure}[htp]
 \centering
 \includegraphics[width=2.9in]{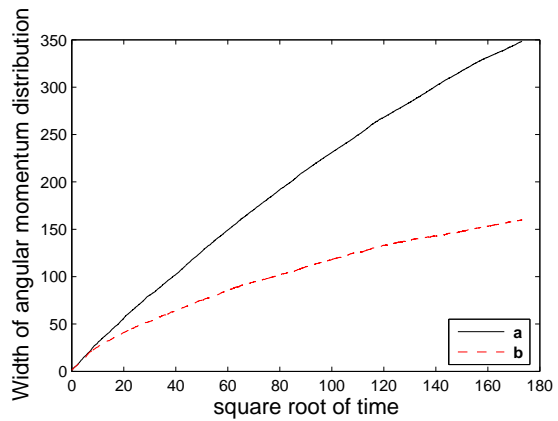}
 \caption{(Color online) Quasi-diffusive regime.\ {\bf a}: $\lambda_{1}= 0.5, \lambda_{2}= 0.5, \lambda_{3}=3.0, \lambda_{4}=3.0$.
 {\bf b}: $\lambda_{1}= 0.25, \:\lambda_{2}= 0.25, \:\lambda_{3}= 3.0, \:\lambda_{4}=3.0$. }
 \label{fig:5}
 \end{figure}

The evolution of the quantum system yields a momentum-space wave function, $\Psi(p_{1},p_{2})$,
as a function of time, for each set of the Hamiltonian parameters.
From it, we obtain the two-rotor momentum probability distribution, $|\Psi(p_{1},p_{2})|^2$,
and calculate the widths (standard deviation) of the single-rotor momentum distributions,
\begin{equation}
\ S = \sqrt{\left < \left(\hat{p}-\left<\hat{p}\right>\right)^{2} \right>}
\label{width}
\end{equation}
as a function of time (measured in kick numbers).
These are plotted as a function of the square root of time, rather than time, 
in Figures \ref{fig:3}, \ref{fig:4}, and \ref{fig:5} because this choice of variable
is effective in showing both short and long time behavior on the same scale.
This plot would yield a straight line for normal diffusion.

We observed three kinds of behavior of the momentum width as a function of time.
\begin{description}
\item[Localized:] The width of the momentum distribution saturates in time,
  as illustrated in Fig.\ref{fig:3}.
\item[Quasi-diffusive:] The width of the momentum distribution increases with time,
  but the growth rate diminishes, as illustrated in Fig.\ref{fig:5}.
\item[Intermediate:] The width of the momentum distribution may exhibit a complex
  oscillatory behavior, as illustrated in Fig.\ref{fig:4}.
\end{description}

Tables \ref{tableX1} and \ref{tableX2} show the occurence of the three behaviors 
as a function of the coupling strengths $\lambda_3$ and $\lambda_4$, for fixed values of the
kicking strengths $\lambda_1$ and $\lambda_2$. 
For vanishing kicking strengths, $\lambda_{1}=\lambda_{2}=0$, all states with non-zero coupling strengths appeared to be localized.

\begin{table}[h]
\caption{Behavior of the angular momentum distribution width for fixed kicking parameters $\lambda_{1}=\lambda_{2}=0.50$. {\bf D}: Quasi-diffusive regime.  {\bf I}: Intermediate regime.  {\bf L}: Localized regime. }
\label{tableX1}
\begin{center}
\begin{tabular}{|c|c|c|c|c|c|c|c|} 
\hline
 $\lambda_{4}$ &0.0 &0.5 &1.0 &1.5 &2.0 &2.5 &3.0 \\
 \hline
$\lambda_{3}$&&&&&&&\\
 0.0 &L &L &L &L &L &L &L\\
 \hline
 0.5 &L &L & L&L &L &I &I\\
 \hline
 1.0 &L &L &L &L &I &I &I\\
 \hline
 1.5 &L &L &L &I &I &I &D\\
 \hline
 2.0 &L &L &I &I &I &D &D\\
 \hline
 2.5 &L &L &I &I &D &D &D\\
 \hline
 3.0 &L &I &I &D &D &D &D\\
\hline
\end{tabular}
\end{center}
\end{table}
\begin{table}[h]
\caption{Behavior of the angular momentum distribution width for fixed kicking parameters $\lambda_{1}=\lambda_{2}=0.25$. {\bf D}: Quasi-diffusive regime.  {\bf I}: Intermediate regime.  {\bf L}: Localized regime. }
\label{tableX2}
\begin{center}
\begin{tabular}{|c|c|c|c|c|c|c|c|} 
\hline
 $\lambda_{4}$ &0.0 &0.5 &1.0 &1.5 &2.0 &2.5 &3.0 \\
 \hline
$\lambda_{3}$&&&&&&&\\
 0.0 &L &L &L &L &L &L &L\\
 \hline
 0.5 &L &L &L &L &L &L &L\\
 \hline
 1.0 &L &L &L &L &L &L &L\\
 \hline
 1.5 &L &L &L &L &L &L &L\\
 \hline
 2.0 &L &L &L &L &I &I &I\\
 \hline
 2.5 &L &L &L &I &I &I &D\\
 \hline
 3.0 &L &L &L &I &I &D &D\\
\hline
\end{tabular}
\end{center}
\end{table}

\begin{figure}[htp]
 \centering
\includegraphics[width=2.9in]{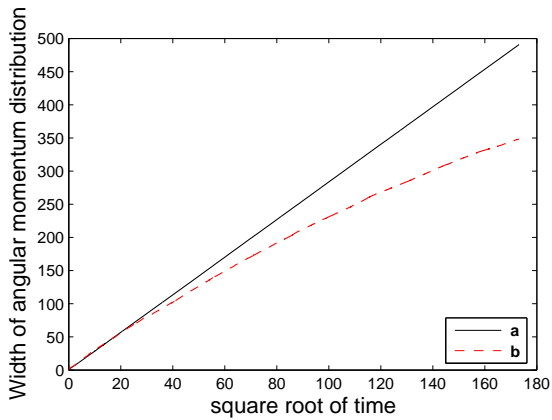}
 \caption{(Color online) Width of angular momentum distribution at early times. \  {\bf a}: classical rotor. 
  \ {\bf b}: quantum rotor.
  $\lambda_{1}= 0.5, \:\lambda_{2}= 0.5,\: \lambda_{3}=3.0, \:\lambda_{4}=3.0$.}
 \label{fig:6}
 \end{figure}

The quasi-diffusive behavior occurs for the largest values of the couplings
between the rotors.
Figure \ref{fig:6} shows that at the earliest times, the quantum rotor behaves diffusively,
as does the classical rotor, but at later times the quantum diffusion rate decreases.
It is not clear whether at very long times the quantum rotor will continue to behave diffusively 
but with a reduced diffusion coefficient, or whether it will eventually saturate at some large 
localization scale.  We shall return to this question in the next section.

\begin{figure}[htp]
 \centering
\includegraphics[width=3.0in]{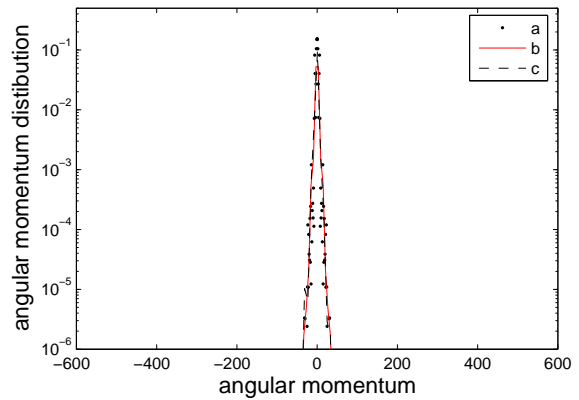}
\caption{(Color online) Localized regime: angular momentum distribution for rotor \#1 along the section $p_2=0$. 
  $\lambda_{1}= 0.5, \:\lambda_{2}= 0.5, \:\lambda_{3}= 0.0, \:\lambda_{4}=3.0$. 
  {\bf a}: at t=1000, {\bf b}: at t=5000, {\bf c}: at t=30000 steps. }
 \label{fig:7}
 \end{figure}

\begin{figure}[htp]
 \centering
 \includegraphics[width=3.0in]{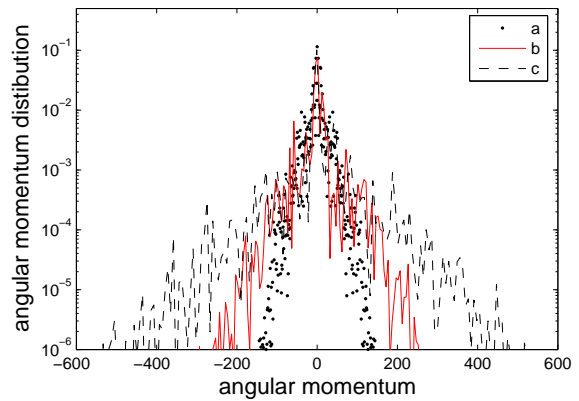}
 \caption{(Color online) Quasi-diffusive regime: angular momentum distribution for rotor \#1 along the section $p_2=0$.
  $\lambda_{1}= 0.5, \:\lambda_{2}= 0.5, \:\lambda_{3}= 3.0,\: \lambda_{4}=3.0$.
  {\bf a}: at t=1000, {\bf b}: at t=5000, {\bf c}: at t=30000 steps. }
 \label{fig:8}
 \end{figure}

The two localization criteria -- exponential fall-off of the angular-momentum distribution, 
and saturation of the momentum-distribution widths -- are consistent with each other.
For states that are classified as ``Localized" in the tables,  we have verified that after 
sufficient time has elapsed, the momentum distribution falls off exponentially with a rate 
that is independent of time. 
For the ``Quasi-diffusive" states, the momentum distribution continues to spread. 
This is illustrated in Figures \ref{fig:7} and \ref{fig:8}, repectively, where we plot the
angular momentum distribution for $p_1$ along the section $p_2=0$.

\section{Scaling Theory and the Quasi-diffusive Regime}

In the Quasi-diffusive regime, the computations do not clearly determine
whether the motion will remain diffusive at very long times, or whether it will become
localized at some scale that is greater than the grid size of the computation.
In this section we apply the scaling theory of Anderson localization (see~\cite{Lee} for a review)
to answer this question.

 In its usual context, the phenomenon of Anderson localization refers to a charged particle 
moving in a disordered medium. 
Scaling theory considers the behavior of the conductance $g$ (not the conductivity) 
as a function of the system size $L$.
The mean-free-path $l$ may be defined as the distance beyond which the phase of the wave
function is essentially randomized by the scattering.  
No phase coherence exists between regions separated by more than $l$, and scaling
behavior is possible only for $L$ greater than $l$.
The conductance $g_0 = g(l)$ at this cutoff length acts as a measure of the disorder 
in the system.

It is useful to distinguish the regimes of strong and weak disorder.
Strong disorder gives rise to exponentially localized states, and to a non-ohmic conductance
that scales as $g(L) \propto \exp(-L/\zeta)$, where $\zeta$ is a localization length that
is generally greater than $l$, but may approach $l$ in extreme cases.
In $d=1$ dimension all states exhibit this {\it strong localization}, while in $d=3$ dimensions
this strong localization occurs only for sufficiently strong disorder.
The case of $d=2$ dimensions may exhibit strong localization, but it can also exhibit
a 
{\it marginal localization} for weak disorder  due to the coupling of the two rotors, for which $g(L)$ differs from ohmic conduction by a
weakly varying correction proportional to $\log(L)$.
Weak disorder usually leads to ohmic conductance, however $d=2$ is an exceptional case.

Abrahams {\it et~ al}~\cite{Abrahams}  have argued that the 
logarithmic derivative of the conductance  $g$ with respect to 
sample size $L$ is a function $\beta(g)$ of the conductance alone,
\begin{align}
\frac{d \log g}{d \log L}=\beta(g)
\end{align}
The form of $\beta(g)$ for ohmic conduction in $d$ dimensions is 
\begin{align}
\beta(g) = (d-2)
\end{align}
For strong localization, it takes the form 
\begin{align}
\beta(g) = \log \left(\frac{g}{g_{0}}\right)
\end{align}
There exists a critical value for $g_{0}$ at which the large scale behavior changes from 
conducting  to localized.

For weak disorder, it is possible to use perturbation theory to calculate the corrections 
to ohmic conduction, which yields
\begin{align} 
\label{beta-gen}
\beta(g) = (d-2) - \frac{a}{g}
\end{align}   
where $a$ is some constant for the system.
It is clear that $d=2$ is a special case. 

We can now increase the size of the system $L$ ($L>l$). 
From (\ref{beta-gen}) for $d=2$, we have
\begin{align}
\frac{d \log g}{d \log L} = -\frac{a}{g}
\label{betaeq}
\end{align} 
Integrating this with respect to $L$ from the lower bound $l$ yields
\begin{align}
g(L) = g_{0} - a\, \log \left [ \frac{L}{l}\right]
\label{glg0}
\end{align}
The localization length, $\zeta$, is defined as the length scale $L$ for which  
 $g(L)=0$, that is
\begin{align}
\zeta = l \, \exp \left[ \frac{g_{0}}{a}\right]
\label{zeta1}
\end{align} 
In transport theory for weak scattering~\cite{Lee}, we have
\begin{align}
g_{0}\propto l
\label{gproptol}
\end{align}
so we can rewrite (\ref{zeta1}) as
\begin{align}
\zeta=l \, e^{b \, l}
\label{zeta2}
\end{align}
where $b$ is a constant for the system. 

Recall that the kicked rotor system can be mapped onto an Anderson-type lattice~\cite{Fishman}. 
If the phase of the rotor's wave function is randomized between two kicks, 
then the average displacement in momentum after the first kick
can be regarded as the counterpart of the mean-free-path in the Anderson lattice.
This should be true for the large values of the rotor coupling parameters, 
for which we obtained quasi-diffusive behavior.
The average displacement in momentum after the first kick is
 \begin{align}
 l^{2} = \sum_{r_{1}, r_{2} = - \infty}^{\infty} (r_{1}^{2}+ r_{2}^{2}) \left| U_{{\it 0},\b{r}} \right|^{2}   
 \label{defl}
 \end{align}
where
\begin{eqnarray}
U_{0,\b{r}}:=\left<0,0\right| \hat{U} \left|r_{1},r_{2} \right>, \;\;\;\;\b{r} = (r_{1},r_{2})
\end{eqnarray}
and $\hat{U}$ is the Floquet operator defined in (\ref{U}). 
The matrix elements are in the angular-momentum basis.
If, in the quasi-diffusive regime, the state is actually localized on some
large scale, then we may expect the scaling theory of Anderson localization to apply.

The initial state for our numerical calculations is a Gaussian, effectively localized in a 
finite portion of the infinite angular-momentum lattice. 
The support of the wave function grows with time.
At any finite time, the effect of lattice sites far outside the wave function's support will be 
negligible, and it would not make any difference to the dynamics if the wave function 
were instead located on a finite lattice of a size not less than the width of the state.  
So it is reasonable to regard the state width $S$ (or some multiple of it) as being the
analog of the sample size $L$ in the Anderson lattice.

It is also reasonable to assume that the diffusion in angular-momentum space of the rotor will
contain the same kind of information as does the diffusion of charged particles on the 
Anderson lattice, and that they will scale similarly with $S$ or $L$, respectively.
According to the Einstein relation~\cite{eins}, the diffusion coefficient is proportional to 
the mobility, and hence to the conductivity.  
But in $d=2$ dimensions, the conductance $g$ scales the same
way with size as does the conductivity.
Thus the diffusion coefficient $D$ of the rotor should scale the same as does the conductance
$g$ of the lattice.

The measure of the angular-momentum state width of the rotor at time $t$ is taken to be  
 $\ S = \sqrt{\left < \left(\hat{p}-\left<\hat{p}\right>\right)^{2} \right>}$. 
By analogy with the classical equation (\ref{cl-dif}), we define the 
diffusion coefficient $D$ of the kicked rotor to be the slope of the line 
relating $S^2$ to $t$. 
If the relationship is not a straight line, as is the case for Fig.(\ref{fig:5}) and (\ref{fig:6})
(which, however show $S$ vs $\sqrt t$),
then we can define a time-dependent diffusion coefficient as the local slope of the curve
of $S^2(t)$ vs $t$.

In the spirit of the scaling hypothesis, we assume that the time-dependent diffusion 
coefficient $D$ at any time $t$ is a function only of $S$ at that time, 
that is to say, $D = D(S(t))$.
In analogy with (\ref{betaeq}), we write
\begin{equation}
\frac{d \log D}{d \log S}=-\frac{a}{D}
\end{equation}
with $a$ being some unknown constant for the system.

In the kicked-rotor model, all quantities change only at the discrete time steps $t_{n}$.  
The state width at time $t_{n}$ is $S_{n}$. 
From the definition of $D_{n}=D(t_{n})$ as the slope of the curve of $S^2$ vs $t$,
we have 
\begin{equation}
\label{s2dif1}
S_{n+1}^{2}-S_{n}^{2} = D_{n} \Delta T
\end{equation}
 where  $\Delta T = t_{n+1}-t_{n}$. 
According to the scaling assumption, we should have a relation of the form 
\begin{equation}
D_{n} =  c \,l - a\log \left[ \frac{S_{n}}{l}\right]
\end{equation}
by analogy with (\ref{glg0}), with two as yet undetermined parameters $c$ and $a$.  
Hence we obtain
\begin{align}
S_{n+1}^{2}-S_{n} ^{2}= \Delta T  \left(c\,l-a \log \left[ \frac{S_{n}}{l}\right]\right)
\label{s2dif2}
\end{align}
As in the original scaling argument that introduced the localization length $\zeta$ in the 
Anderson lattice, we can define the saturation length $\Lambda$ 
as the value of $S_{n}$ that sets the right hand side of the above equation to zero. 
This yields a result analogous to (\ref{zeta2}),
\begin{equation}
\label{lscale}
\Lambda=l \, e^{bl}
\end{equation}
with $b=c/a$.
Since the corresponding parameter $b$ was a constant for the Anderson lattice,
we expect that here $b$ will be approximately constant in the Quasi-diffusive regime 
of the rotor if the condition of  
marginal localization for weak disorder is valid.

\begin{table}
\caption{Numerical values of $l$, $b$ and $\Lambda$ [See (\ref{defl}) and (\ref{lscale})] for each run with  $\lambda_{1}=\lambda_{2}=  0.50 $}
\label{tableY3}
\begin{center}
\begin{tabular}{|c|}
\hline
$l$\\
\hline
\end{tabular}
\end{center}
\begin{center}
\begin{tabular}{|c|c|c|c|c|c|c|c|} 
\hline
 $\lambda_{4}$ &1.0 &1.5 &2.0 &2.5 &3.0 \\
 \hline
$\lambda_{3}$&&&&&\\
 1.0 & & & & &3.57\\
 \hline
 1.5& & & &3.37 &3.86\\
 \hline
 2.0 & & &3.20 &3.67 &4.15\\
 \hline
 2.5 & &3.06 & 3.52&3.98&4.46\\
 \hline
 3.0 & 2.96 & 3.39&3.84 &4.30 & 4.77\\
\hline
\end{tabular}
\end{center}
\begin{center}
\begin{tabular}{|c|}
\hline
$b$\\
\hline
\end{tabular}
\end{center}
\begin{center}
\begin{tabular}{|c|c|c|c|c|c|c|c|} 
\hline
 $\lambda_{4}$ &1.0 &1.5 &2.0 &2.5 &3.0 \\
 \hline
$\lambda_{3}$&&&&&\\
 1.0 & & & & &0.87\\
 \hline
 1.5& & & &0.86&1.01\\
 \hline
 2.0 & & &0.84 &1.02 &1.19\\
 \hline
 2.5 & &0.97 & 1.00&1.11&1.11\\
 \hline
 3.0 & 0.82 & 0.93&1.14 &1.22& 1.16\\
\hline
\end{tabular}
\end{center}
\begin{center}
\begin{tabular}{|c|}
\hline
$\Lambda$\\
\hline
\end{tabular}
\end{center}
\begin{center}
\begin{tabular}{|c|c|c|c|c|c|c|c|} 
\hline
 $\lambda_{4}$ &1.0&1.5&2.0&2.5 &3.0 \\
 \hline
$\lambda_{3}$&&&&&\\
 1.0&&&&&80\\
 \hline
 1.5 &&&&61&190\\
 \hline
 2.0 &&&47&155&579\\
 \hline
 2.5 &&60&115&330&630\\
 \hline
 3.0 &34&79&306&816& 1207\\
 \hline
\end{tabular}
\end{center}
\end{table}
\begin{table}
\caption{Numerical values of $l$, $b$ and $\Lambda$ for each run with  $\lambda_{1}=\lambda_{2}=  0.25$}
\label{tableY31}
\begin{center}
\begin{tabular}{|c|}
\hline
$l$\\
\hline
\end{tabular}
\end{center}
\begin{center}
\begin{tabular}{|c|c|c|c|c|c|c|c|} 
\hline
 $\lambda_{4}$ &1.0&1.5&2.0&2.5 &3.0 \\
 \hline
$\lambda_{3}$&&&&&\\
 1.0&&&&&3.54\\
 \hline
 1.5 &&&&3.34&3.83\\
 \hline
 2.0 &&&3.17& 3.64&4.13\\
 \hline
 2.5 &&3.03&3.49&3.96&4.43\\
 \hline
 3.0 &2.92&3.36&3.81&4.27& 4.75\\
\hline
\end{tabular}
\end{center}
\begin{center}
\begin{tabular}{|c|}
\hline
$b$\\
\hline
\end{tabular}
\end{center}
\begin{center}
\begin{tabular}{|c|c|c|c|c|c|c|c|} 
\hline
 $\lambda_{4}$ &1.0 & 1.5 &2.0  &2.5 &3.0 \\
 \hline
$\lambda_{3}$&&&&&\\
1.0 &&&&&0.60\\
 \hline
  1.5 &&&&0.65&0.71\\
 \hline
 2.0 &&&0.57&0.67& 0.80\\
 \hline
 2.5 &&0.57&0.69&0.79& 0.86\\
 \hline
 3.0 & 0.58 &0.67&0.77 &0.83&0.79\\
\hline
\end{tabular}
\end{center}
\begin{center}
\begin{tabular}{|c|}
\hline
$\Lambda$\\
\hline
\end{tabular}
\end{center}
\begin{center}
\begin{tabular}{|c|c|c|c|c|c|c|c|} 
\hline
 $\lambda_{4}$ &1.0&1.5&2.0&2.5 &3.0 \\
 \hline
$\lambda_{3}$&&&&&\\
 1.0&&&&&30\\
 \hline
 1.5 &&&&29&58\\
 \hline
 2.0 &&&19&42&112\\
 \hline
 2.5 &&17&39&90&201\\
 \hline
 3.0 &16&32&72&149& 202\\
 \hline
\end{tabular}
\end{center}
\end{table}

We now wish to test the scaling theory by comparison with the results of numerical simulation.
In the previous section, we computed $S^2$ as a function of $t$,
for a range of the kicking and coupling parameters.  These results are now regarded
as data.  
It should be noted that, although the diffision coefficient $D = d(S^2)/dt$ plays a
fundamental role in developing the scaling theory, it is not necessary to compute that derivative
numerically.  Only the data $S^2$ vs $t$ is needed for the computation.

To carry out the scaling-theory calculation, we first compute
 $l$ from (\ref{defl}). The matrix elements in (\ref{defl}) can be obtained by running 
the numerical simulation for one kicking step from an initial state that is localized
at $p_{1}=p_{2}=0$.  These values of $l$ are contained in Tables \ref{tableY3} and \ref{tableY31}.
Next we use the recurrence relation (\ref{s2dif2}) to obtain the scaling theory values 
for $S_n^2$, which we denote as $\tilde{S}_n^2$, using $\tilde{S}_0^2=l^2$ as the starting value. 
The parameters $c$ and $a$ cannot be determined directly, so we must do the recursion calculation
for several values of $c$ and $a$, and determine the best values of these two parameters
by minimizing the sum of squared differences between the scaling-theory values $\tilde{S}_n^2$
and the previously computed data for $S^2(t_n)$.

The choice of $\Delta T$ in (\ref{s2dif1}) and (\ref{s2dif2}) must be made so that $\tilde S^2(t)$
varies nearly linearly with $t$ during the interval $\Delta T$.
Taking $\Delta T=1$ kicking step has the disadvantage of yielding a curve whose slope might  
fluctuate considerably from step to step.  It is better to choose a larger value of $\Delta T$
so as to smooth out such short-time fluctuations, but of course $\Delta T$ must not be so
large as to cover a range over which the curve is appreciably nonlinear.
We found the choice of $\Delta T=300$  kicking steps to be a good compromise, but the results
are insensitive (in the first decimal place) to the precise value that was chosen. 

The least-squares fitted values of $a$ and $c$ were used to calculate the  parameter $b=c/a$.
The values for $b$ and the corresponding saturation length $\Lambda$, from (\ref{lscale}),
are listed in Tables \ref{tableY3} and \ref{tableY31}.  
The values of $b$ seem to become nearly constant as we move to the lower right of the tables,
away from the transition zone between  the Localized and Quasi-diffisive regions shown in
Tables \ref{tableX1} and \ref{tableX2}.  
This is what we should expect if the scaling hypothesis is valid, and our 
Quasi-diffusive regime is actually a regime of  
marginal localization for weak disorder.
In contrast to the nearly constant parameter $b$, the localization length $\Lambda$ 
varies over orders of magnitude in these ranges of kicking and coupling strengths.  
Our results are consistent with 
marginal localization for weak disorder on the scale of $\Lambda$.

\section{Conclusions}

This paper was motivated by an apparent contradiction between the conclusions of two 
published papers~\cite{Doron,Ikeda1} on very similar models of coupled kicked rotors. 
One group found quantum localization, while the other reported diffusive behavior.
It was unclear whether the differing conclusions were due to the different criteria for
localization used by the two groups, or whether they reflected fundamental differences
between the two models, or some other reason.
By studying a more general model, of which those two are special cases, we were able to
resolve the apparent contradiction.  The two models do, indeed, belong to the same generic 
class and have similar behavior.  When the two criteria for localization are applied to the
same model in the same range of parameters, they always agree.  
The differences between the results reported
by the two groups were primarily due to their use of very different interaction strengths and kicking intervals.

 Beyond resolving this discrepancy in the literature, our study of two
coupled kicked rotors with two different coupling interactions has revealed some 
interesting systematic behavior as the kicking and coupling strength parameters are varied.
The classical model exhibits diffusive motion.
An analytic formula for the dependence of the classical diffusion coefficient on the kicking 
and coupling strength parameters was derived, by means of a random-phase approximation.
The numerically computed diffusion coefficient oscillates about this approximate formula,
much as occurs for uncoupled rotors.
As the couplings between the rotors become large, the behavior of the rotors is determined
mainly by the sum of the two coupling parameters, $\lambda_3 + \lambda_4$.  
This generalization holds, at least qualitiatively, for the quantum rotors too.

The quantum mechanical model shows classical diffusion at short times, but at long times
the diffusion is limited by a kind of quantum localization.
For zero coupling strength, the model reduces to two independent kicked rotors,
each of which exhibits the well-established quantum localization that is the analog of 
Anderson localization in a one-dimensional random lattice.  
This single rotor localization behavior  
 persists for a moderate range of couplings between the two rotors.

As the coupling strength increases further, the system passes through a complex transition zone,
before entering a quasi-diffusive regime for strong coupling.  In this quasi-diffusive regime,
the mean squared width of the momentum distribution increases with time, but the slope of the
curve (the local diffusion coefficient) gradually declines.
Within practical limits of grid size and simulation time, it was not possible to determine
directly whether the motion is bounded on a much larger scale or grows without bound.
We found that the 
marginal localization for weak disorder  due to the coupling of the two rotors fits well to this quasi-diffusive motion,
and thereby we conclude that the state is very probably localized on a larger scale 
than our practical computational grid.
We were also able to estimate this large localization scale, which varies smoothly with
the coupling strength parameters.
In the intermediate region between strong and weak localization, the state-width
oscillates in a complicated way, but its envelope always lies beneath that found in 
the regime of weak localization. 

Thus we have shown that a system of two coupled kicked rotors exhibits a complex set of
behaviors, including a transition from strong to weak localization. 
This would be expected from the formal mapping of the two-rotor system onto a two-dimensional
Anderson lattice.  However, this correspondence is not trivial, since the theory of 
Anderson localization assumes a {\it random} lattice potential, whereas the effective potential
generated by the kicked-rotor system is not random, but merely quasi-periodic with 
incommensurate frequencies.  Thus the existence of strong and weak localization in the rotor
system is not guaranteed by the theory of localization in random lattices.

However, we disagree with one conclusion of Adachi {\it et al}~\cite{Ikeda1},
who claimed that, for sufficiently large coupling strength, the coupled 
quantum rotors exhibit normal diffusion.  
Their results were obtained for a special set of parameters, 
for which the dimensionless kicking interval (\ref{alphas}) is exactly 
commensurate with $\pi$. Such special cases give rise to periodic lattices, 
for which the quantum states are not localized, regardless of the kicking and 
coupling strengths. These highly special cases do not correspond to 
typical physical situations. 
It is more realistic to study the generic non-commensurate case. 
We believe that, in the generic case, the two coupled quantum rotors exhibit 
weak localization on a long time scale.

Finally, we mention a possible connection between this model and experiment. 
Experimental realization of systems like the one considered in the present paper are nowadays intesnly explored. The experimental observation of 2-dimensional optical localization have been reported~\cite{Nat,PRL}, where 
the  localized results were shown to match with the predictions of scaling theory. 
\newpage
 \begin{acknowledgments}
BT acknowledges support from Alberta's Informatics Circle of Research Excellence (iCORE), Canada's National Sciences and Engineering research Council (NSERC), and the Canadian Centre of Excellence for Mathematics of Information Technology and Complex Systems (MITACS QIP and  MITACS US ARO).
 \end{acknowledgments}

\end{document}